\documentstyle[twoside,fleqn,espcrc2]{article}

\def\ov{\overline}
\def\s2{\frac{1}{\sqrt2}}

\def\beq{\begin{equation}}
\def\eeq{\end{equation}}
\def\beqa{\begin{eqnarray}}
\def\eeqa{\end{eqnarray}}
\def\times{\otimes}
\def\IR{\relax{\rm I\kern-.18em R}}
\def\IP{\relax{\rm I\kern-.18em P}}
\def\IC{\relax\hbox{\kern.25em$\inbar\kern-.3em{\rm C}$}}

\def\cp#1{\relax\ifmmode {\IP\kern-2pt{}_{#1}}\else
$\IP\kern-2pt{}_{#1}$\fi}

\def\im{\hbox{Im\,}}
\def\re{\hbox{Re\,}}

\def\IZ{{\bf Z}}

\def\beq{\begin{equation}}
\def\eeq{\end{equation}}

\def\Dsl{\hbox{/\kern-.6700em\it D}} % D slash

\def\s{\sigma}

\def\dsl{\hbox{/\kern-.5300em$\partial$}}
\def\IR{\relax{\rm I\kern-.18em R}}
\font\cmss=cmss10 \font\cmsss=cmss10 at 7pt
\def\IZ{\relax\ifmmode\mathchoice
{\hbox{\cmss Z\kern-.4em Z}}{\hbox{\cmss Z\kern-.4em Z}}
{\lower.9pt\hbox{\cmsss Z\kern-.4em Z}}
{\lower1.2pt\hbox{\cmsss Z\kern-.4em Z}}\else{\cmss Z\kern-.4em Z}\fi}

\def\inbar{\vrule height1.5ex width.4pt depth0pt}
\def\IC{\relax\thinspace\hbox{$\inbar\kern-.3em{\rm C}$}}

\newcommand{\AmS}{{\protect\the\textfont2
  A\kern-.1667em\lower.5ex\hbox{M}\kern-.125emS}}

\title{Superstring Phenomenology: An Overview}

\author{Fernando Quevedo\address{Instituto de F\'{\i}sica,\\
Universidad Aut\'onoma de M\'exico\\
Apartado Postal 20-364, 01000\\
M\'exico D.F., M\'exico}%
        \thanks{Email:fernando@ft.ifisicacu.unam.mx}}
        
\begin{document}

\begin{abstract}
The different aspects of superstring phenomenology before and after 1995
are briefly reviewed.

\vspace{12pt}

%{\vbox{\noindent
%\begin{center}
%{ \em To the hundreds of thousands%
% of guatemalan citizens who 
%sacrificed their lives  during more than }
%%
%
%{\em three decades of civil  war,  especially to those I had the privilege 
% to know and love,
%and now I  miss.}
%\end{center}%}}}
\end{abstract}

\maketitle

\section{INTRODUCTION}

String phenomenology has been under intense study for  more than twelve years
already. The main subtopics we can distinguish on this field are:
 model independent
results, model building, low-energy effective actions and supersymmetry
 breaking. Here we will briefly mention the current status of each of these
 subtopics, comparing the situation before 1995 and after 1995.

\section{SUPERSTRING PHENOMENOLOGY BEFORE 1995}

Before 1995 there were five consistent string theories,
all existing in ten dimensions, namely, type I open strings
with gauge symmetry $SO(32)$,
type IIA and IIB closed strings and the two closed heterotic strings
with gauge symmetries $E_8\times E_8$ and $SO(32)$.
Out of these five theories, the one that attracted most of the attention
was the heterotic $E_8\times E_8$ theory because it was the most 
promising for phenomenology: upon compactification to 4D it gives rise to 
chiral $N=1$ supersymmetric models with an observable sector, coming from the
first $E_8$ which contains the standard model symmetry $SU(3)\times SU(2)
\times U(1)$ and several families of matter fields. The second $E_8$ gives
rise to a hidden sector, which fits perfectly with the  attempts of
supersymmetric model building prior to string theory, there
a hidden sector was proposed that breaks supersymmetry at an intermediate scale
$\sim 10^{12}$ GeV and gravity plays the role of messenger of supersymmetry breaking to the observable sector which feels the breaking of supersymmetry 
 at the electroweak scale $\sim 10^{2}$ GeV.

The other four string theories were much less interesting,
there was even a no-go theorem proving the impossibility to obtain the standard
 model out of type II theories. 

At that time,
 we could only extract perturbative information
in discussing phenomenological aspects of string theory
because the theory itself was   formulated  only  at the perturbative level. 
Therefore nonperturbative effects, such as those 
induced by the stringy versions of 
instantons, monopoles etc, were not under control.
The issues of supersymmetry breaking and dilaton potentials  needed
to be addressed at the nonperturbative level and the only 
possibility was to consider field theoretical nonperturbative effects
such as the condensation of gauginos in the hidden sector.

The tools used to study string phenomenology were threefold:
\begin{description}
\item{(i)}Conformal field theory (CFT) with all its technology was heavily used to build
`exact' solutions of string theory giving rise to 4D string models
with the phenomenologically desired properties, also CFT was applied to explicitly
compute the important couplings such as gauge and Yukawa couplings
directly from string theory  and moreover, to extract general properties of string
vacua which are independent of the model, although are perturbative in nature.

\item{(ii)}Powerful topological tecniques were used in the geometrical interpretation
of string compactifications in terms of compact 6D Calabi-Yau manifolds, in order to construct compactifications, compute (cubic) Yukawa couplings and scalar
 kinetic terms and study the
`moduli space' of these solutions.

\item{(iii)}Finally, some `macroscopic' techniques were used in order to extract 
information about the 4D effective action. Mostly based on general symmetry 
arguments. This allowed to prove remarkable results such as the 
nonrenormalization of the superpotential and of the gauge kinetic function 
beyond one-loop.

\end{description}

The combination of these three different approaches allows us to have a general understanding of model building, effective actions and model independent results in string perturbation theory, as well as supersymmetry breaking from nonperturbative
field theoretical effects. We will briefly discuss these subjects here,
for more complete treatments including the relevant references we refer the reader
 to \cite{gsw}, \cite{bert1}, \cite{dine}. In particular the next subsections
are based on the long review  \cite{yo}\  and I will refer to this paper
not only for a longer exposition but also  for a
detailed guide to the  references. See also \cite{nathkane}.

\subsection{Model Building}

The geometrical  approach  of 4D string model building can be described 
{\it \`a la} Kaluza-Klein  in the sense of viewing the 10D spacetime
as a product of 4D Minkowski spacetime and a small 6D compact manifold.
This manifold is constrained by the requirement of having $N=1$ supersymmetry
to be a Calabi-Yau manifold, which is a 3D complex space with holonomy 
$SU(3)$. There are many  such string `vacua',
first because of the many possible Calabi-Yau manifolds, which are not
completely classified yet, but  also because, for each compactification, we still
have the freedom of choosing the  embedding in the gauge degrees of freedom
(of the heterotic string for instance). The particular embedding identifying
the spin connection and the gauge connection gives rise to a $(2,2)$
supersymmetric  CFT, where the entries refer to the number of supersymmetries for
 left and right moving string modes, respectively. 
The generic embedding however has only 
$(0,2)$ supersymmetry. The $(0,2)$ Calabi-Yau compactifications are not yet
well understood and most of the work on this area has been done on the $(2,2)$
models.

The simplest way to construct $(0,2)$ models    is by constructing explicit 
CFTs such as orbifold compactifications. An orbifold is a twisted torus
that happens to be a particular singular limit of a Calabi-Yau manifold.
The advantage of working with orbifolds is that, similar to flat space or torus compactifications they correspond to free CFTs, their only complication relies in the
choice of boundary conditions which is understood.
 Similarly, by either bosonizing or fermionizing all the coordinates of the extra
space (the orbifold) and choosing different boundary conditions for each of the
fields we arrive at the `fermionic' and `lattice' constructions of 4D models.
Furthermore, there are more complicated constructions based on coset CFTs
and known as Gepner-Kazama-Suzuki models, they attracted substantial attention
 especially after the realization that the $(2,2)$ versions of these models were
also particular points of a Calabi-Yau manifold despite their original 
 nongeometrical formulation.

Out of all these different formulations of models we learned that there is a 
twofold degeneracy of string models. One is discrete, given by say topologically different Calabi-Yau compactifications with different numbers of generations
and antigenerations of quarks and leptons (for $(2,2)$ Calabi-Yau's these numbers are topological, in particular the net number of generations minus antigenerations is
given by the Euler number of the manifold.).   The second degeneracy is continuous,
{\it i.e.} for each compactification, there are a number of free parameters or 
`moduli' which can be varied freely, these moduli can describe the size and shape
of the internal space and in the 4D effective  theory they correspond to
 massless scalar
fields `$T$', having vanishing potential. Therefore the continuous degeneracy 
is given by  the  moduli space of the $T$ fields defined by the
existence of  flat directions.
Due to the existence of non-renormalization theorems, the flatness of the potential
stays to all orders in perturbation theory.

This enormous degeneracy limits the predictive power of the theory.
We might expect that the continuous degeneracy may be lifted by the existence of 
non-perturbative string effects which could create a nonvanishing scalar
 field potential, but the discrete degeneracy seems more serious.
Before 1995 it seemed impossible to be able to compare two different
 compactifications
differing for instance on the total number of generations since,
 having topological origin, it was impossible to deform one space into.%

Another piece of  information we can extract about this degeneracy of string models
is by taking a phenomenological approach and search among all of these models
those which resemble the standard model of particle physics
with   3 families of quarks and leptons and standard model
gauge group $SU(3)\times SU(2)\times U(1)$.

Some quasi-realistic models have been
studied with much detail using this approach. This
includes models with three families and standard model gauge group
$SU(3)\times SU(2)\times U(1)$, $SU(4)\times SO(4)$
 as well as a version of
$SU(5)\times U(1)$ known as flipped $SU(5)$
\cite{yo}, \cite{faraggi}. 
All of
these  models also reproduce many  of the nice features of the
standard model (proton stability, doublet-triplet splitting,
pattern of fermion masses etc.). Nevertheless, as in the case of orbifolds, there
is not a totally realistic model yet.
In particular there is no model yet with just the spectrum of the
supersymmetric standard model.

  An unsolved issue
in most of these models is the fact that after breaking to the standard model
group, the number of extra light doublets is not well determined
since their mass is not generally protected by symmetries and
the actual calculation of the remaining superpotential, after
solving the $D$ and $F$ flatness conditions, is done only 
up to operators of a given dimension. The presence of extra
light doublets in orbifold models was understood in terms
of the selection rules established for those models, but they are
precisely the main source of problems for those models also, because
of their contribution to the running couplings and therefore
to the string unification scale and Weinberg's angle. The extra
light doublets is
then an obstacle for obtaining realistic models in all the different
approaches studied so far. Furthermore, without understanding the breaking of
 supersymmetry, we cannot confront directly these models with physics.

The more recent progress in this area was the successfull construction of 4D
string GUTs, {\it i.e.} string models with simple Grand Unified groups
(such as $E_6$ and $SO(10)$) and an adjoint representation that can  break  the
group to the standard model. The presence of adjoints requires special string
 constructions named `higher level Kac-Moody models' which are a particularly 
difficult-to- construct subclass of the $(0,2)$ string models discussed above.
These constructions are expected to share the phenomenological virtues of 
supersymmetric
GUTs,  especially in what refers to the unification of the gauge couplings.
The models obtained so far have not been studied in great detail but they still
 suffer from some difficulties, such as fine tunning and the possible survival of
 extra
states at low energies that could modify the running of the gauge couplings
\cite{kakush}.
Therefore it is still an open question to construct a fully realistic string model.

\subsection{Effective Actions}

The general Lagrangian coupling $N=1$ supergravity 
to gauge and chiral multiplets
 depends on three arbitrary functions of the chiral multiplets:

\begin{description}
\item{(1)}The K\"ahler potential $K(z,\ov z)$ which is a {\it real}
function. It determines the kinetic terms of the chiral 
fields 
\beq
{\cal L}_{kin}=K_{z\ov z}\partial_\mu z\partial^\mu \ov z
\eeq
with $K_{z\ov z}\equiv \partial^2 K/\partial z\partial\ov z$.
% with subindices
%reflecting derivatives with respect to the corresponding variable.
$K$ is called K\"ahler potential because the manifold
of the scalar fields $z$ is K\"ahler, with metric $K_{z\ov z}$.
 
\item{(2)}The superpotential $W(z)$ which is a {\it holomorphic}
function of the chiral multiplets (it does not depend on
$\ov z$)
%\footnote{Actually, W is  a section of a line bundle .}.
 $W$ determines the Yukawa couplings as well
as the $F$-term part of the scalar potential $V_F$:
\beq
%_F(z,\ov z)=
e^{K/M_p^2}\left\{ D_zW\, K^{-1}_{z\ov z}\ov{D_zW}-3\frac{|W|^2}{M_p^2}\right\},
\eeq
with $D_zW\equiv W_z+WK_z/M_p^2$. Here and in what follows, the internal
indices labelling different chiral multiplets $z_i$ are not
explicitly written.

\item{(3)}The gauge kinetic function $f_{ab}(z)$ which 
is also {\it holomorphic}. It determines the gauge kinetic terms
\beq
{\cal L}_{g}= \re f_{ab} F_{\mu\nu}^a\, F^{\mu\nu\, b}
+\im f_{ab} F_{\mu\nu}^a\, \tilde F^{\mu\nu\, b}
\eeq
it also contributes to gaugino masses and the gauge
part of the scalar potential.
\beqa
V_D &=&\left(\re f^{-1}\right)_{ab}\left(K_z, T^a \, z\right)\, \left(K_{\ov z}, T^b\, \ov z\right)\nonumber\\
V &=& V_F+V_D
\eeqa
 
\end{description}

These three functions are arbitrary for a generic $N=1$ supersymmetric model,
but in string theory we should be able to compute them for each model.
General non-renormalization theorems can be applied to the holomorphic functions
$W$ and $f$. The main reason is that the dilaton field $\phi$ is always present in 
string models and its vev is the loop counting parameter. In $N=1$ 4D models
 it joins an axion-like field $a$ having a Peccei-Quinn symmetry
 $a\rightarrow a +$ constant, to  
form a complex (chiral) multiplet $S=\phi+i a$, because of the
existence of the Peccei-Quinn symmetry $W$ cannot depend on $a$ and, because
$a$ only appears through $S$,
 then $W$ cannot depend on $S$ and so it does not depend
on the loop-counting-parameter, therefore it is not renormalized.
 A similar argument applies to $f$ beyond one loop.
The Peccei-Quinn symmetry is usually broken by nonperturbative effects therefore
they will contribute to corrections to $W$ and $f$.
 For $K$ there are no simplifications and very little is known beyond some
 tree level calculations.

The general structure of these functions of matter multiplets $Q$, moduli $T$ and
dilaton $S$ is the following:
\begin{eqnarray}
W(S,T,Q) = W_{tree}(T,Q)+W_{np}(S,T,Q)\nonumber\\
f(S,T,Q) = S+f_{1-loop}(T,Q)+f_{np}(S,T,Q)\nonumber\\
K(z,\ov z)=(K_{tree}+K_{1-loop}+\cdots)+ K_{np}
\end{eqnarray}
Therefore we can see that the lack of control on the perturbative corrections to $K$
is the main source of ignorance of the full perturbative 4D effective actions.

\subsection{Model Independent Results}

Let us  recapitulate in this section  what we can say about string models 
which is independent of the model.
This is the closest we can get to string predictions and
help us in approaching general questions, differentiating the
generic issues from those of a particular model.
Since the full nonperturbative formulation of string theory is not yet
available, we have to content ourselves mostly with 
predictions of string perturbation theory, assuming that
the corresponding string model is given by a CFT.

\begin{description}
\item{(i)}
String models
{\it predict} the existence of gravity and gauge interactions.
This is a point that cannot be overemphasized since it is the
first theory that makes those fundamental predictions for interactions
we experience in the every day life.

\item{(ii)}
The dimension of spacetime is dynamical and $D\leq 10$ raising the
hope that eventually we could explain if a 4D spacetime
is in some way  special, although at present. Also
the rank of the gauge group is bounded $r\leq 22$
%\footnote{Both of the last two statements have been recently
%modified at the non-perturbative level, 
%by the studies of strong-weak coupling duality
%symmetries (for a recent review see \cite{joe}\ ). 
%In particular there is some evidence for the
%existence of an 11D theory from which all the different strings
%could emerge. There is also evidence for the appearance
%of nonperturbative gauge groups that can raise the rank beyond $r=22$
%\cite{witteninst}.} 

\item{(iii)}There are other fields which 
survive at low energies: charged matter fields $Q$, candidates to be
 basic building  blocks of matter but also the dilaton field $S$ and the moduli
$T$. We have to mention  that, although as yet there
is no 4D model without moduli fields, there is not
a general theorem implying their existence. In that sense the dilaton is the most generic modulus field, with a flat potential
in perturbation theory.

\item{(iv)}There is only one arbitrary parameter 
$\alpha'$ fixed to be close to the Planck scale
$M_p$. All other parameters
of the effective action are determined by expectation values of fields
such as the dilaton and the moduli. In particular the gauge coupling is
given at tree level by the vev of $S$.

\item{(v)} The existence of spacetime supersymmetry is needed for consistency,
although $N=1$ is selected for phenomenological reasons.
There is a general requirement for a CFT to lead to $N=1$ spacetime supersymmetry:
 It has to have $(0,2)$ supersymmetry in the wordlsheet
(2D) (plus  a quantization condition on the
charges of the $U(1)$ group mixing the two supersymmetries).

\item{(vi)} There are no global internal symmetries in 4D string
models, besides the already mentioned Peccei-Quinn symmetry of
the $S$ field and some accidental global symmetries
(like baryon and lepton numbers in the standard model).
%This is a very strong result derived by showing that if there is
%an internal symmetry, the properties of CFT's imply that there should
%be a vector field in the spectrum with the properties of the gauge field of that
%symmetry. This is consistent with similar claims about the
%nonexistence of global symmetries in gravity, due for instance to 
%wormhole effects.
 This puts very strong constraints to string models
compared with standard field theory models.

\item{(vii)}There are generically some {\it discrete} symmetries
in string models. Some infinite dimensional
such as $T$-duality which in the simplest version takes the form of an
$SL(2,\IZ)$ transformation
\beq
T\rightarrow\, \frac{a\,T+b}{c\, T+d}
\eeq
with $a,b,c,d$ integers satisfying $ad-bc=1$
There are also finite dimensional discrete symmetries, such 
as those inherited from the twist  defining  orbifold constructions,
  which are seen as
discrete gauge symmetries in the 4D effective theory.
 These can in principle be useful for
model building, hierarchy of masses etc. There are however some couplings
that vanish in string theory and {\it cannot} be explained in terms of symmetries
 of the effective 4D theory, these are called `string miracles'
since from the 4D point of view they seem to break the criterium for naturalness.
 $T$-duality symmetries restrict 
very much the form of the effective action and quantities such as 
Yukawa couplings have to be modular forms of a given duality group.
These symmetries are valid to all orders in string perturbation theory
and are thought to be also preserved by nonperturbative effects.
Matter fields $Q^I$ are assigned special quantum numbers,
the modular weights $n$, according to their transformation properties
 under the duality group. For a $SL(2,\IZ)^m$ group for instance we have:
\beq
Q^I\rightarrow (ic_l\, T_l+d_l)^{n^l_I}\, Q^I,\quad l=1,\cdots ,m.
\eeq
Since fermions transform nontrivially under these symmetries, there
may be `duality anomalies' which have to be cancelled for consistency.
This imposes strong constraints on the possible spectrum of the
corresponding string model.

\item{(viii)}
There is unification without the need of a GUT. If 
the gauge group is a direct product of several groups we have
for the heterotic string:
%we 
%assume the standard model group $SU(3)\times SU(2)\times U(1)$
%at low energies, the corresponding gauge couplings satisfy:
\beq
k_1\, g_1^2=k_2\, g_2^2=\cdots=\frac{8\pi}{\alpha'}G_{N}\equiv
g^2_{string}.
%\frac{k_1}{g_1^2}=\frac{k_2}{g_2^2}=\frac{k_3}{g_3^2}=\frac{1}{g_{grav}^2}
\eeq
Where $k_i$ are special stringy constants known as the Kac-Moody levels of the corresponding
gauge groups (for the standard model groups it is usually assumed that
$k_2=k_3=1, k_1=5/3$), $g_i$ are the gauge couplings and $G_N$ is Newtons constant.
We can see there is a difference with standard GUTs in field theory
for which we compute the unification scale by finding the
point where the different string couplings meet.
In heterotic string theory, the unification scale is given in terms of the
string coupling $g_{string}$ and the 
Planck scale.\footnote{For type I strings 
the gravitational and gauge couplings are independent, so 
we have the freedom to adjust the unification scale with experiment
as in usual GUTs}
 More precisely:
$M_{string}\sim 5.27\times 10^{17} g_{string}$ Gev.
For $g_{string}\sim {\cal O}(1)$ this shows a discrepancy with the
`observed' value of the unification scale given by the
experiments $M_{GUT}\sim 2\times 10^{16}$ GeV. Also 
the Weinberg angle gives $\sin^2\theta_W=0.218$ differing from the
experimental value of $\sin^2\theta_W=0.233\pm0.0008$.
Therefore the string `predictions' are
very close to the experimental value, which is encouraging, but
differ by  several standard deviations from it. This is the 
string unification problem. The situation looks much better
for simple GUT's which have good agreement with experiment.
Several ideas have been proposed to cure this problem, including 
large values of threshold corrections, intermediate scales, extra
particles, changing the values of Kac-Moody levels etc, with 
no compelling solution yet (for recent discussions of this
issue see for example   the review of K. Dienes in ref.
 \cite{keith}.).  

\item{(ix)} There are usually fractionally charged particles in 4D string models.
In fact it can be shown that we cannot have simultaneously
$k_2=k_3=3k_1/5 =1$ in the standard model and only integer charged
particles, because if that is the case the standard model 
gauge group would be enhanced to a full level-one
$SU(5)$.  This `problem' can be evaded in models where  the
fractionally charged particles are heavy string states, it has
also been proposed that those particles could confine at intermediate energies and 
be unobservable.

\item{(x)}
There are  `anomalous' $U(1)$ groups in most
of the models, but  there is also a counterterm in the action cancelling the anomaly and  generating a Fayet-Iliopoulos kind of term :
\beq
\frac{1}{S+\ov S}\left\vert\frac{Tr q^a}{48\pi^2}\frac{1}{(S+\ov S)^2}+\sum{q_I^a |Q_I| ^2}\right\vert^2,
\eeq
where $q_I^a$ are the anomalous charges of the scalar fields $Q_I$.
This term is responsable to break the would be anomalous group by fixing the 
value of a combination of the matter fields $Q_I$, breaking 
the would be anomalous $U(1)$ and usually other gauge groups,
 but not supersymmetry (although this has not been shown in general). 
 A combination of the fields $Q_I$ and the dilaton $S$
still remains massless and plays the role of the new dilaton field.

\end{description}

There are  further model independent results which refer to nonperturbative
string effects and  will be discussed  next.

\subsection{Supersymmetry Breaking}

As we discussed previously there are two main problems of string perturbation theory
namely, the enormous vacuum degeneracy and supersymmetry breaking.
There were two lines of research towards attacking these problems.
\begin{description}
\item{(i)}
To use a particular field-theoretical nonperturbative effect which is the
 condensation of gauginos induced after the asymptotically free hidden sector
 becomes strongly interacting at lower energies ($\sim 10^{12}$ GeV).
\item{(ii)}
To consider the supersymmetry breaking sector as a black box and study its
 possible implications assuming that a combination of the moduli fields $T$ 
and the dilaton $S$ are responsable for breaking supersymmetry,
 but without specifying how.
\end{description}

The second approach has been recently reviewed 
in \cite{bim}\ and I refer the reader to that
review for the general results.
As for gaugino condensation, it was also reviewed in 
\cite{yo},\cite{mirev}, we will only mention here that the combination of
$T$ duality with the inclusion of several condensing groups with matter in the
 hidden sector (very generic in string theory) can give rise to interesting results,namely, supersymmetry is broken at the phenomenologically desired scale
($\sim 10^2$ Gev),  with the moduli $T$ and dilaton $S$ fixed at the 
desired values $T\sim 1$, $S\sim 2$ in Planck scale units.
 This value of $S$ gives the 
expected value of the gauge coupling at the unification scale.
However there were several problems:
\begin{description}
\item{(i)} The cosmological constant is very large and negative.

\item{(ii)} The dilaton potential is runaway and there has to be 
some fine tunning in order  to obtain a nontrivial minimum 
besides the minimum at zero coupling and $S\rightarrow \infty$.

\item{(iii)}
 There are
at least two serious cosmological problems for the gaugino condensation
scenario. First, it was found under very general grounds, that 
it was not possible to get inflation with the type of dilaton potentials
obtained from gaugino condensation \cite{bs}. Second is the so-called
`cosmological moduli problem' which applies to any (non-renormalizble)
hidden sector  scenario including gaugino condensation
\cite{dccqr}. In this case,
it can be shown that if the same effect that 
fixes the vev's of the moduli, also breaks supersymmetry,
then: the moduli and dilaton fields acquire masses
of the electroweak scale ($\sim 10^2$ GeV) after supersymmetry breaking
\cite{dccqr}.
Therefore if  stable, they overclose the universe, if
 unstable, they  destroy  nucleosynthesis by their 
late decay, since they only have gravitational strength interactions.
At present there is no satisfactory explanation of this problem and it stands
as one of the unsolved generic problems of string phenomenology.

\end{description}

The runaway behaviour of the dilaton has been argued to be a generic
problem for string models \cite{dineseib}. The reason for this is that 
being $S$ the string coupling, we know that for $S\rightarrow\infty$
the theory becomes free and then the scalar potential has to vanish.
This was used in \cite{dineseib}, to argue that strings have to be strongly 
coupled in order to develop another  minimum, unless some parameters conspire to 
fix $S$ at weak coupling. This argument has been revised recently
in \cite{bmmq}. There it was argued that even if the scalar potential vanishes
at $S\rightarrow\infty$ in the full theory, in an effective theory, after
 integrating out some of the massive fields the remaining potential for $S$ could 
blow up for $S\rightarrow\infty$.
A simple example of this is the $\lambda\phi^4$ theory since upon
minimizing the potential
$V=-\frac{1}{2}m^2\phi^2+\frac{1}{4}g^2\phi^4$ we find $\ov\phi=\pm m/g$
and substituting back into $V$ gives $V(\ov\phi)=-m^4/4g^2$ which blows up as the coupling $g\rightarrow 0$. A particular mechanism to achieve this could be 
to consider nonasymptotically free models which appear very often in string
 theories. If the potential blows up at $S\rightarrow\infty$, then the cosmological
problems discussed in \cite{bs} are not necessarily there.
Therefore this is an interesting way of fixing the dilaton that deserves further
investigation.

All this has been said using mostly nonperturbative field theoretical effects but
 we know that there should also be stringy nonperturbative effects that could
 play a role in fixing the dilaton and breaking supersymmetry.
 In general we should always consider the two types
of nonperturbative effects: stringy (at the Planck scale) and
field theoretical (like gaugino condensation). Four different scenarios
can be considered depending on which class of mechanism solves
each of the two problems:lifting the vacuum degeneracy and breaking
supersymmetry. 

 For breaking supersymmetry at low energies, we expect that
a field theoretical effect should be dominant in order
to generate the hierarchy of scales (it is hard to believe that
a nonperturbative effect at the Planck scale could generate
the Weinberg-Salam scale). 
We are then left with two preferred scenarios:
either the dominant nonperturbative effects 
solve both problems simultaneously, or there is a `two steps' scenario 
in which stringy effects dominate to lift  vacuum degeneracy 
and field theory effects dominate to break supersymmetry.
The first scenario has been the only one considered so far,
it includes gaugino condensation.
The main reason this is the only scenario considered so far is that we
 can control field theoretical
nonperturbative effects but not the stringy. In this scenario,
 independent of the particular mechanism, we have to face the cosmological
 moduli problem.
In the two steps scenario stringy effects only lift the vacuum degeneracy and
supersymmetry may be broken dynamically by field theoretical effects,
such as discussed in a whole session at this conference
\cite{dsb}. We have to conclude that there is no prefer scenario yet.

\section{SUPERSTRING PHENOMENOLOGY AFTER  1995}
The recent progress in understanding nonperturbative issues of string theory
\cite{filq2},
\cite{joe} has necessarily strong impact on the phenomenological questions,
we are only starting to explore these implications which can be 
sumarized as follows.
\begin{description}
\item{(i){$\ $}}{\it Unification of theories:}
We mentioned in the introduction that there are five consistent superstring theories and each has thousands or millions of different vacua.
It is now believed that the five string theories are related by 
strong-weak coupling dualities and furthermore, they appear 
to be different limits of a single underlying fundamental theory,
probably in 11D, the $M$ theory
(probably related with membranes or higher dimensional objects such as five-branes), 
which is yet to be constructed.
If this is true it may solve the arbitrariness in the number 
of fundamental string theories by deriving them from a single
theory.
\item{(ii){$\ $}}{\it Unification of vacua (?):} Recent work
based on comparison of string compactifications with the Seiberg-Witten
theory, has lead to the conclusion that many and probably all
Calabi-Yau compactifications are connected. Then it seems
that not only the five different theories are unified, but also
all the vacua of these theories could
also be unified: since, if they are all connected,  we can foresee a
 mechanism that lifts the
degeneracy and select one point in the web
of compactifications, something it could not have been done before 
because they were thought to be disconnected vacua.
 These  transitions occur in singular points of the
corresponding moduli space where a particular state
(massless black hole) or even an infinite tower of states (tensionless strings) become massless.
They were partially understood  for $N>1$ compactifications,
but recently extensions  to the phenomenologically interesting
$N=1$ case have been found
\cite{ks},  implying  for instance that models with different
number of families would belong to the same moduli space reducing in some sense the 
discrete degeneracy problem to the level of the continuous degeneracy problem
  and so we may expect that probably
one particular  number of families  could
 eventually be selected dynamically.
\item{(iii)}{\it Nonperturbative vacua:}
The fact that the strong coupling regime of a given string theory
would simply be the weak coupling regime of another string theory
would be very dissapointing since that means that the problems present at 
weak coupling would remain at strong coupling.
Fortunately this is not the case. For instance, the strong coupling limit of
the $E_8\times E_8$ string is believed to be given by $M$ theory compactified in 
the orbifold
$S^1/\IZ_2$ which is just a one dimensional interval.
$M$ theory contains elementary membranes and their magnetic dual, 
$5-$branes. The membranes can end at each of the two $10 D$ ends
of the interval (fixed points) which are $9$-branes and 
generate an $E_8$ symmetry at each end. The distance between the two 
$9$-branes $\rho$ is proportional to the heterotic coupling
and when this is very small the two $E_8$'s collapse to a single
$10D$ point which is the heterotic string. For any finite coupling 
the membrane  is a cylinder between the two $9$-branes with 
heterotic strings at the intersection. This reproduces the standard
perturbative spectrum of heterotic strings.
The new ingredient comes mostly from the $5$-branes which 
for the $E_8\times E_8$ case, carry two-index antisymmetric tensors,
therefore introducing more  than one of these fields in the spectrum
after compactifications (in the $SO(32)$ version they may lead to extra vector 
fields depending on the compactification).
In perturbative heterotic string there was a single antisymmetric tensor
$B_{\mu\nu}$ that we saw is dual to an axion field.
The appearance of several of those fields in the spectrum shows clearly that the corresponding vacuum is nonperturbative and may 
eventually create more possibilities for using these axion fields
for solving the strong CP problem in string theory.
There is even a model with zero tensor fields. This may be relevant because
$B_{\mu\nu}$ is a  supersymmetric partner  of the dilaton and
having a model without antisymmetric tensors  would mean that somehow the dilaton was fixed, lifting the corresponding degeneracy, and acquired a mass
(avoiding the cosmological moduli problem)!
Furthermore for compact spaces with nontrivial $4$-cycles, the
corresponding $5$-branes could wrap around those cycles
giving rise to another string (different of course from the one obtained
from the membrane). These nonperturbative strings will generically have their own nonperturbative gauge group, therefore enhancing the maximum rank required in perturbation theory
\cite{witteninst}\  (the world record seems to be right now a group of rank of order
 $10^5$!
\cite{philip}).
The physical relevance of the nonperturbative gauge fields is yet to be explored.
\item{(iv){$\ $}}{\it Scales in M theory}: It is interesting to analyze the different scales present in a $4D$ model built from $M$-theory.
There are three relevant scales:  the $11D$ Planck scale $\kappa$,
the length of the interval $\rho$ and the overall volume of the 
compactified $6D$ space $V$. In the $11D$ theory, the gauge and gravitational couplings
can be written as:
\begin{eqnarray}
L=-{1\over 2\kappa^2}\int_{M^{11}}
 d^{11}x \sqrt g R -\nonumber\\
 \sum_i
{1\over 8\pi
(4\pi \kappa^2)^{2/3}}\int_{M^{10}_i}d^{10}x\sqrt g {\rm tr} F_i^2.
\end{eqnarray}
Where $M^{11}$ is the $11D$ space (bulk) and $M_i^{10}$, $i=1,2$ are
the two $10D$ $9$-branes at each end of the interval.
We can see that after compactification, the $4D$ Newton constant
and gauge couplings are given by $G_N={\kappa^2\over{16 \pi^2 V\rho}}$
and $\alpha_{GUT}={\left(4\pi \kappa^2\right)^{2/3}\over
{2V}}$. Notice that now $M_{GUT}^2=V^{-1/3}={\alpha_{GUT}\over 8\pi^2 G_N^{2/3}\rho}$, since we have an extra parameter, $\rho$, we can get $M_{GUT}\sim 10^{16} GeV$
 by setting $\rho^{-1}\sim 10^{12-14} Gev$
something we could not have done in perturbative heterotic strings.
 This has been used by Witten to claim that it may be possible to solve the string unification problem
by tunning the extra parameter as in standard GUTs \cite{wittenn}. 
We then get the following picture: at large distances the universe looks $4D$ 
at energy  scales between $10^{12-14} GeV$ and $10^{16} Gev$ it looks $5D$ and 
at higher scales (smaller distances)  it looks $11D$.
This new intermediate scale ($\rho$) may play an interesting role for other 
phenomenological and cosmological questions. There is a complication that 
for $\rho^{-1}\leq 10^{15} GeV$ the gauge coupling of one of the gauge groups 
blows up, this has been argued by Witten that could put a bound on
Newton's constant on a generic model. There are some specific models which
avoid this problem which makes them more attractive.
Also, the
process of gaugino condensation can be reanalyzed in this picture
\cite{hor}, \cite{ignat}.
 A single condensate in the hidden $E_8$  $9$-brane, does not break supersymmetry in its vecinity nor in the $5D$ bulk but due to a topological obstruction it can break
supersymmetry in the observable sector \cite{hor}. Note that in this picture the standard model lives at one of the `end of the world' branes while gravity and the moduli fields live in the $5D$ bulk. The possible physical consequences of this new picture are only starting to be 
explored \cite{wittenn}.
\item{(v){$\ $}}{\it Nonperturbative superpotential:}
 It is  quite remarkable that recently Witten and others have been able to extract information about  superpotentials derived from stringy nonperturbative effects
\cite{wittsup}.  At the moment there have been found three classes
of results, depending on the compactification:$W=0$, $W\sim e^{-\Phi}$, $W=$ a modular form.
Here $\Phi$ is one of the moduli fields.
The first case is interesting because it means there are compactifications for
which the nonperturbative superpotential vanishes so the only source of
 superpotential could be strong coupling infrared effects such as gaugino
 condensation making the field theoretical discussion above more
relevant. The second case gives the standard runaway behaviour of the scalar potential and the third possibility is a realization of the kind of duality invariant potentials  proposed in the past
\cite{flst},\cite{filq2}, in this case there are nontrivial minima and it
 is yet to be studied in detail whether supersymmetry could be broken,
 in particular these models seem suitable for a realization of the two steps
 scenario alluded to before. We hope more progress will be made in this direction
 which is addressing the main problem of superstring phenomenology 
from a nonperturbative formulation.

\item{(vi)}{\it Stringy $e^{-1/g}$ effects}: Some time ago, Shenker proposed that in string theory,
there would appear nonperturbative effects of the form $e^{-1/g}$ on top of the standard field theoretical effects of the form $e^{-1/g^2}$. These have been 
argued to correct the K\"ahler potential and contribute to the dilaton potential in such a way that the dilaton can be fixed even with a single exponential in $W$
\cite{unog}. Recently, these effects were explicitly computed for the heterotic string for a particular compactification \cite{silv}.

\end{description}

We can see that many of the results from string perturbation theory are modified
by the nonperturbative information obtained so far.  Some of the other results 
are expected to be modified or need revision, for instance the nonexistence of
 global symmetries
was proved using CFT techniques which are explicitly perturbative, it is expected that being string theory a theory of gravity, global symmetries will not be allowed
(as usually found studying black holes and wormholes), but a general nonpertubative proof 
is not available yet. Also, the main problems such as supersymmetry breaking, 
 are still open  which  is a good motivation to work on this field.


\begin{thebibliography}{9}
\bibitem{gsw}M. Green, J. Schwarz and E. Witten, 
{\em Superstring Theory, volumes 1,2},
Cambridge University Press (1987);
M. Kaku, {\em String Theory} Springer-Verlag (1988);
D. L\"ust and S. Theisen, {\em Lectures in String Theory},
Springer Lecture Notes in Physics, Vol. 346 (1989);
P. Ginsparg, {\em Les Houches Lectures} Elsevier (1989);
J. Polchinski, {\em Les Houches Lectures}, hep-th/9411028
and book to appear.

\bibitem{bert1} B. Schellekens, {\em Superstring Construction},
North-Holland (1989).

\bibitem{dine} M. Dine, {\em String Theory in Four Dimensions},
North-Holland (1988).


\bibitem{yo}F. Quevedo, hep-th/9603074 and references therein.

\bibitem{nathkane}P. Nath and G. Kane contributions to this conference.

\bibitem{faraggi}For a recent review see: A.Farggi, hep-ph/9707311.


\bibitem{kakush}Z. Kakushadze, contribution at this conference.

\bibitem{keith}
K. Dienes, hep
-th/9602045 and references therein.

\bibitem{bim}A. Brignole, L. Ib\'a\~nez and C. Mu\~noz, hep-ph/9707209.

\bibitem{mirev}F. Quevedo, hep-th/9511131.

\bibitem{bs}{R. Brustein, P. Steinhardt, {\em Phys. Lett.} B302 (1993) 196.}

\bibitem{dccqr}T. Banks, D. Kaplan, A. Nelson, {\em Phys. Rev.} D49 (1994) 779B;
 de Carlos, J.A. Casas, F. Quevedo and E. Roulet, {\em Phys. Lett.} B318 (1993) 447.


\bibitem{dineseib}M. Dine and N. Seiberg, {\em Phys. Lett.} B162 (1985) 299.

\bibitem{bmmq}C.P. Burgess, A. de la Macorra, I. Maksymyk and F. Quevedo,
hep-th/9707062.

\bibitem{dsb}See the contributions of Nelson, Dine, Poppitz and Murayama
at this conference.

\bibitem{filq2}A. Font, L.E. Ib\'a\~nez, D. L\"ust, F.
Quevedo, {\em Phys. Lett.} 249B (1990) 35;
S.-J. Rey, {\em Phys. Rev.} D43 (1991) 526;
C. Hull and 
P.K. Townsend, {\em Nucl. Phys. } B438, (1995) 409;
E. Witten, {\em Nucl. Phys.} B443 (1995) 85.


\bibitem{joe} For  recent reviews  see: J. Polchinski, hep-th/9607050;
J. Schwarz, hep-th/9607201, P.K. Townsend, hep-th/9612121, C. Vafa,
hep-th/9702201.

\bibitem{ks}{S. Kachru and E. Silverstein, hep-th/9704185;
G. Aldazabal, A. Font, L.Ib\'a\~nez, A. Uranga and G. Violero, hep-th/9706158.}

\bibitem{witteninst} E. Witten, hep-th/9507121.

\bibitem{philip}P. Candelas, E. Perevalov and G. Rajesh, hep-th/9704097.

\bibitem{wittenn}E. Witten, hep-th/9602070;
 T. Banks and M. Dine, hep-th/9605136.

\bibitem{hor}P. Ho\v rava, hep-th/9608019.

\bibitem{ignat} I. Antoniadis and E. Dudas,  contributions to this conference.

\bibitem{wittsup} E. Witten, {\em Nucl. Phys.} B474 (1996) 343. 
See also the contribution of D. L\"ust to this conference.

\bibitem{flst}S. Ferrara, D. L\"ust, A. Shapere and S.Theisen,
{\em Phys.Lett.} B225 (1989) 363;
E. Chun, J. Mas and H.P. NIlles, {\em Phys.Lett.} 233B (1989) 141.

%\bibitem{bdine}{T. Banks, M. Dine, {\em Phys. Rev.} D50 (1994) 7454.}




\bibitem{unog}Y.-Y. Wu, contribution to this conference.

\bibitem{silv}{E. Silverstein, hep-th/9611195.}



\end{thebibliography}
\end{document}